\documentclass[english,russian,11pt,twoside]{article}
\usepackage{amssymb,amsmath}
\usepackage[russian]{babel}
\usepackage[cp866]{inputenc}
\begin{document}
\title{ГАМИЛЬТОНОВА ФОРМА УРАВНЕНИЙ МАКСВЕЛЛА И ЕЕ ОБОБЩЕННЫЕ РЕШЕНИЯ }
\author{\bf{Людмила А. Алексеева }}
\date{}
\maketitle \centerline{\textit {Институт математики МОН
РК,\,ул.Пушкина, 125,Алма-Ата,050010 Казахстан} }
\centerline{alexeeva@math.kz } \vspace{10mm}

\begin{abstract}
Построено   векторное уравнение для комплексного A-поля,
эквивалентное системе уравнений Максвелла для электромагнитных
полей. Рассмотрены сильные ударные волны, получены условия на
фронтах для скачков напряженностей $A,E,H$ и обобщенные законы
сохранения. Доказана поперечность ударных волн. Построены тензор
Грина, обобщенные решения модифицированного уравнения для
нестационарных, стационарных и монохроматических  полей и
обобщенное решение задачи Коши. Доказана единственность
классического решения задачи Коши, в том числе при наличии ударных
волн.
\end{abstract}

Система уравнений Максвелла для описания электромагнитных (ЭМ) полей обладает симметрией относительно
векторов напряженностей электрического $(E)$ и магнитного $(H)$ полей. Этот факт хорошо известен в
электродинамике при построении ее решений. Решение краевых задач для электрических полей часто можно
использовать для построения решений аналогичных задач для магнитных полей простой заменой
$E\rightleftarrows H$ и констант электрической и магнитной проницаемости $(-\varepsilon\leftrightharpoons\mu)$
и наоборот. Симметрию этих уравнений нарушает только условие отсутствия магнитных зарядов и соответственно
магнитных токов.

Здесь мы эти условия снимаем и строим одно комплексное дифференциальное уравнение для комплексного
трехмерного векторного A-поля, эквивалентное системе уравнений Максвелла. Для исследования этого уравнения,
называемого \textit{гамильтоновой формой }уравнения Максвелла [1], используем методы теории обобщенных функций,
а именно подход для решения нестационарных краевых задач электродинамики, разработанный в [2]. Этот подход
позволяет строить и исследовать сравнительно просто разрывные решения уравнения со скачком не только
производных, но и самих искомых функций. Последнее имеет важное значение для изучения нестационарных волновых
процессов типа ударных волн.

 {\bf 1. Классические уравнения Максвелла.} Система уравнений
Максвелла для однородной изотропной среды имеет следующий вид :
\begin{equation}\label{(1.1)}
 - \varepsilon\,\partial _t E + rot\,H = j^E ,
 \quad\mu\, \partial _t H + rot\,E = j^H.
\end{equation}
Здесь электрические и магнитные проницаемости $ \varepsilon ,\;\mu
$ -положительные константы, $ E,H $- напряженности электрического
и магнитного поля, $ j^E (x,t),j^H (x,t) $ - плотности
электрических и магнитных токов, $ x = (\,x_1 ,\,x_2 ,\,x_3) ,\, t
\ge 0. $

Согласно полной системе уравнений Максвелла
\begin{equation}\label{(1.2)}
\varepsilon\, div\, E = \rho ^E ,\quad - \mu\, div\, H = \rho ^H
,\quad \rho ^H = 0,\quad j^H  = 0,
\end{equation}
$ \rho ^E ,\rho ^H $-- объемные плотности электрических и
магнитных зарядов.  Далее последние два  условия отсутствия
магнитных зарядов и токов  снимаем.

Взяв дивергенцию в (\ref{(1.1)}), получим в дифференциальной форме

\textit{Закон сохранения зарядов}:
\begin{equation}\label{(1.3)}
\partial _t \rho ^E  + div\,j^E= 0,\quad\partial _t \rho ^H+div\, j^H  =
0.
\end{equation}

Если уравнения (\ref{(1.1)}) скалярно умножить на $E$ и $H$
соответственно и вычесть, получим в дифференциальной форме

\textit{ Закон сохранения энергии}:
\begin{equation}\label{(1.4)}
\partial _t W + div\,P = (j_{^{} }^H ,H) - (j_{}^E ,E) .
\end{equation}
Здесь $ P = E \times H $ - вектор Пойнтинга, $ W =0,5 \left(
{\varepsilon \left\| E \right\|^2  + \mu \left\| H \right\|^2 }
\right) $ - плотность энергии ЭМ поля.

При заданных токах уравнения (\ref{(1.1)}) достаточны для
определения ЭМ поля $(E,H)$. В этом случае равенства (\ref{(1.2)})
служат для определения заряда, а законы сохранения выполняются
тождественно.

{\bf 2. Гамильтонова форма уравнений Максвелла.} Введем комплексный
вектор напряженности:

\begin{equation}\label{(2.1)}
A = \sqrt \varepsilon \, E + i\,\sqrt \mu \, H .
\end{equation}
Тогда систему (\ref{(1.1)}) можно записать в  виде одного
векторного уравнения:
\begin{equation}\label{(2.2)}
 - c^{ - 1} \partial _t A -i\,rot\,A = j,\quad c = 1/\sqrt {\varepsilon \mu, }
\end{equation}
где $c$- скорость ЭМ волн, а  комплексные токи определяются
выражением $ j = \sqrt {\mu }\, j^E  - i\,\sqrt \varepsilon \,
j^H. $

О п р е д е л е н и е. Назовем комплексным зарядом А-поля величину
\begin{equation}\label{(2.3)}
\rho  =c ^{ - 1} div\,A  =  \sqrt \mu \, \rho ^E  - i\,\sqrt
\varepsilon \,\rho ^H .
\end{equation}
Если взять дивергенцию (\ref{(2.2)}), то получим для комплексного
заряда тот же

\textit{Закон сохранения заряда}:
\begin{equation}\label{(2.5)}
\partial _t \rho  + div\,j = 0.
\end{equation}
Заметим, что энергия ЭМ  поля  при такой записи определяется через
модуль комплексного вектора поля: $ W = 0,5\left\| A \right\|^2  =
0,5(A,A^* )$ ,  а вектор Пойнтинга
$
P =- 0,5\,ic \,A \times A^* $ . Здесь  $ A^*
$-комплексно-сопряженное $A$. Последнее следует из равенства:
\begin{equation}\label{(2.7*)}
A\times
A^*=(\sqrt{\varepsilon}\,E+i\,\sqrt{\mu}\,H)\times(\sqrt{\varepsilon}\,E-i\,\sqrt{\mu}\,H)=
2\,\sqrt{\varepsilon\mu}\,i\,E\times H.
\end{equation}

Умножим скалярно (\ref{(2.2)})  на  $cA^* $, сложим это равенство
с комплексно-сопряженным. В результате получим

\textit{Закон сохранения энергии}:
\begin{equation}\label{(2.9)}
\partial _t W + div\,P =  - c\,{\mathop{\rm Re}\nolimits} (j,A^* ).
\end{equation}

Таким образом вместо обычной системы уравнений Максвелла имеем ее
комплексную форму в виде одного векторного уравнения, которую
гораздо удобнее исследовать, т.к. она содержит в два раза меньше
уравнений, зависит только от одной постоянной $c$ - скорости
света, и полностью определяет ЭМ поле, токи и заряды. При этом
законы сохранения заряда и энергии имеют свою обычную
формулировку.

{\bf 3. Ударные волны.} Обозначим  $L(\partial _x ,\partial _t )$-
матричный дифференциальный оператор уравнений (\ref{(2.2)}),
компоненты которого имеют вид:
\[
L_{kj}(\partial_x,\partial_t)  =  - c^{ - 1} \delta _{kj} \partial
_t  + ie_{kjl}
\partial _l ,\quad l,j,k = 1,2,3,
\]
где $\delta _{kj}$-символ Кронекера, $e_{jlk} $- единичный
псевдотензор Леви-Чивита. Здесь и всюду по повторяющимся индексам
суммирование от 1 до 3 (тензорная свертка).

Легко показать, что характеристическое уравнение (\ref{(2.2)}): $
det\{ L_{kj}(\nu,\nu_t) \} = 0$,--  в пространстве $ R^4  = {\rm
\{ }{\bf x} = {\rm (}x,x_4  = ct{\rm )\} } $ приводится к виду:
\begin{equation}\label{(3.1)}
\nu _4 (\nu _4^2  - \nu _{\rm 1}^2  - \nu _2^2  - \nu _3^2 ) = 0.
\end{equation}
Характеристическое уравнение  (\ref{(1.1)}) имеет аналогичный вид
[1]: $ \left( {\nu _4 {\rm (}\nu _4^2  - \nu _{\rm 1}^2  - \nu
_2^2  - \nu _3^2 {\rm )}} \right)^2  = 0.$

В частности, они справедливы при $ \nu _4  = 0 $, что
соответствует поверхности вида $F(x) = const $, которая не
зависит от времени. В окрестности таких поверхностей задача Коши,
согласно теореме С. Ковалевской [3], неразрешима. Т.е. зная  ЭМ
поле на какой-то неподвижной поверхности, невозможно восстановить
его в ее окрестности. Это также говорит о том, что в статической
постановке эти три уравнения не являются независимыми и
недостаточны для решения краевых задач.

Эти уравнения  содержат также конус характеристических нормалей -
классический световой конус: $ \nu _4^2  = \left\| \nu
\right\|^2 ,\quad \left\| \nu  \right\| = \sqrt {\nu _1^2  + \nu
_2^2  + \nu _3^2. } $ Как известно, на характеристических
поверхностях решения и их производные могут терпеть скачки.
Обозначим через $F$ такую поверхность, $ (\nu ,\nu _4 ) = (\nu _1
,\nu _2 ,\nu _3 ,\nu _4 ) $ - нормаль к ней в $R^4 $. В
пространстве координат $R^3 $ ей соответствует волновой фронт
$F_t $, движущийся со скоростью $c$, $n =(n_1,n_2,n_3)$ -
единичный вектор нормали к фронту волны, направленный в сторону
ее движения. Легко видеть, что
\begin{equation}\label{(3.2)}
\nu _4  = - \left\| \nu  \right\|, \quad n_i  = \nu _i /\left\|
\nu \right\|,\quad \;i = 1,2,3.
\end{equation}
Далее для вывода условий на фронтах используем методы теории
обобщенных функций.

Введем пространство обобщенных вектор-функций $ D_{_3 }' (R^4 )  =
\left\{ {\hat f({\bf x}) = (\hat f_1 ,\hat f_2 .\hat f_3 )}
\right\}$ - непрерывных линейных функционалов, определенных на
пространстве бесконечно дифференцируемых финитных функций $ D_3
(R^4 ) = \{\varphi ({\bf x}) =  (\varphi _1 ,...\varphi _3 )\} ,\,
\hat f_k  \in D'(R^4 ) , \,\varphi _k  \in D(R^4 ) $: $ (\hat
f,\,\varphi ) = \sum\limits_{k = 1}^3 {(f_k ,\,\varphi _k
)},\,\,x\in R^4.$ Если $A$ решение (\ref{(2.2)}) с конечным
разрывом на $F$, то в $ D_3' (R^4 )$, согласно правилам
дифференцирования обобщенных функций [2],
\begin{equation}\label{ (3.3)}
\partial _j \hat A = A,_j  + {\rm [}A{\rm ]}_F \nu _j\,
 \delta _F {\rm (}{\bf x}{\rm )},\quad j = 1,...,4.
\end{equation}
Здесь первое слагаемое справа -- классическая производная по $ x_j
,\, \left\| {\bf \nu } \right\| = 1,\, \delta _F $ - простой слой
на $F$, плотность которого определяется через скачок:
$
{\rm [}A{\rm ]}_F  = \mathop {\lim }\limits_{\varepsilon  \to  +
0} \left( {A({\bf x} + \varepsilon {\bf \nu }) - A({\bf x} -
\varepsilon {\bf \nu })} \right)$. В силу непрерывности вне фронта
волны и поскольку поверхность $t=const$ не является
характеристической,
\[
\left[ A \right]_F  = \mathop {\lim }\limits_{\varepsilon  \to  +
0} \left( {A(x + \varepsilon \,n,t) - A(x - \varepsilon \,n,t)}
\right) = A^ +  (x,t) - A^ -  (x,t) = \left[ A \right]_{F_t }.
\]
Операция дифференцирования  позволяет ввести точное определение
понятия    поверхностных зарядов.

О п р е д е л е н и е. Обобщенным зарядом A-поля будем называть
$\hat \rho  =c^{-1} div\hat A.$

Если на поверхности $S \in R^3 $ поле терпит скачок, то $ \hat
\rho =c ^{ - 1} div\,A + c^{ - 1} (n,{\rm [}A{\rm ]}_S )\delta _S
(x)$, где $n$--единичная нормаль к $S$. Первое слагаемое описывает
плотность обычных объемных зарядов, а второе - поверхностных.

О п р е д е л е н и е. Назовем поверхностной дивергенцией вектора
$A$ простой слой $div_S A =(n,\,[A]_S )\,\delta _S  (x)$.

Предполагается, что нормаль $n$ существует почти всюду на $S$, а
скачок $A$ принадлежит классу суммируемых на $S$ функций:
 $[ A]_S\in L_1 (S).$

О п р е д е л е н и е. Поверхностным зарядом A-поля будем называть
выражение вида
$
\hat \rho _S  =c ^{ - 1} (n,{\rm [}A{\rm ]}_S )\delta _S (x) =c ^{
- 1} div_S A.
$

Отметим также , что формула для поверхностного заряда подобна
формуле заряда (\ref{(2.3)}).

О п р е д е л е н и е. Назовем решение уравнений (\ref{(2.2)})
обобщенным, если
\[
(L(\partial _x ,\partial _t )\hat u,\varphi ) \equiv (L_{ik} \hat
u_k ,\varphi _i ) =  - (\hat u_k ,L_{ik} \varphi _i ) = (j_k
,\varphi _k ),\quad \forall \varphi\in D_3(R^4).
\]
Если $A$ классическое решение уравнения (\ref{(2.2)}), разрывное
на $F$, то рассматриваемое как обобщенное, оно удовлетворяет
следующей системе:
\begin{equation}\label{(3.4)}
L_{kj} \,(\partial _x ,\partial _t )A_j  + G_{kj} [A_j ]_F
\;\delta _F  = j_k ,\quad G_{kj}  =  - \delta _{kj} \nu _4  +
ie_{kjm} \nu _m.
\end{equation}
Из (\ref{(3.4)}) следует

Т е о р е м а 3.1. \textit{Для того, чтобы A было обобщенным
решением (\ref{(2.2)}), необходимо выполнение условия}:
\begin{equation}\label{(3.5)}
G_{kj} {\rm [}A_j {\rm ]}_F  = 0,\quad k = 1,2,3.
\end{equation}
Т.е. обобщенное решение может быть разрывным на фронтах. Такие
решения описывают \textit {ударные} ЭМ-волны.

С учетом введенных обозначений равенство (\ref{(3.5)}) на
подвижных волновых фронтах в $R^3 $ приобретает вид, который
представим в следующей теореме.

Т е о р е м а 3.2. \textit{Решение уравнения (\ref{(2.2)}),
непрерывное и дифференцируемое всюду, за исключением волновых
фронтов, на которых выполняются условия (\ref{(3.2)}), является
его обобщенным решением, если
\begin{equation}\label{(3.6)}
\left[ A \right]_{F_t }  = -  i\left[ A \right]_{F_t }  \times m,
\end{equation}
или, для векторов электрической и магнитной напряженностей},-
\begin{equation}\label{ (3.7)}
\varepsilon ^{1/2} \left[ E \right]_{F_t }  = \mu ^{1/2} \left[ H
\right]_{F_t }  \times m,\quad \mu ^{1/2} \left[ H \right]_{F_t }
= \varepsilon ^{1/2} m \times \left[ E \right]_{F_t}.
\end{equation}
Здесь и выше знак "$\times$"обозначает векторное произведение.
Далее скалярное и векторное произведение $a$ и $b$ обозначается
соответственно $(a,b) = a_j b_j,\:[a,b ]:[a,b ]_l = e_{ljk} a_j
b_k $.

Д о к а з а т е л ь с т в о. Из теоремы 3.1 следует:
$
\nu _4 \left[ {A_k } \right]_F  = ie_{kjm} \nu _m \left[ {A_j }
\right]_F.
$
Отсюда, с учетом (\ref{(3.2)}) получим:
$
\left[ {A_k } \right]_F  =  - ie_{kjl} m_l \left[ {A_j }
\right]_F,
$
что соответствует векторной записи формул теоремы.

Из этой теоремы легко выводится ряд полезных следствий.

С л е д с т в и е 1.  На фронтах ударных ЭМ волн $(\left[ A
\right]_{F_t } ,n) = 0 ,$ что для $E$ и $H$ принимает вид
\begin{equation}\label{ (3.9)}
(\left[ E \right]_{F_t } ,n) = 0,\quad(\left[ H \right]_{F_t } ,n)
= 0.
\end{equation}
Т.е. на фронтах ударных волн отсутствуют поверхностные заряды.

С л е д с т в и е 2. Если перед фронтом волны ЭМ поля отсутствуют
$(E^+=0,\, H^+=0)$, то на фронте тройка векторов $E, H, m$ -
взаимно ортогональны. Т.е. вектора $E,H$ лежат в касательной
плоскости к фронту волны, а вектор Пойнтинга $P$ параллелен
волновому вектору. Значит ударные ЭМ волны являются поперечными.

С л е д с т в и е 3. $ \left[ E\right]_{F_t }  = 0\rightleftarrows
\left[ H \right]_{F _t}  = 0$ . Т.е. если электрическое поле
непрерывно, то магнитное тоже непрерывно и наоборот.

Рассмотрим скачок энергии на фронте волны.

Т е о р е м а 3.2. \textit{ На фронте ударной волны}
\begin{equation}\label{(3.10)}
{\rm [}W(x,t){\rm ]}_{F_t }  = c^{ - 1} {\rm (}m,{\rm [}P{\rm
]}_{F_t } {\rm )}.
\end{equation}

Д о к а з а т е л ь с т в о. Умножим (\ref{(3.6)}) скалярно на
комплексно-сопряженное: $ \left[ {A_l } \right]_{F_t } A_l^{*-}  =
- ie_{ljk} \left[ {A_k } \right]_{F_t } A_l^{ *-} m_j.
$
Аналогично, соответствующее комплексно сопряженное выражение
умножим на $A_l^ +$:
$
\left[ {A_l^* } \right]_{F_t } A_l^ +   = ie_{ljk} \left[ {A_{_k
}^* } \right]_{F_t } A_l^ +  m_j.
$
Складывая эти два выражения, с учетом тождества $ {\rm [}ab{\rm ]}
= a^ +  b^ +   - a^ -  b^ -   = a^ +  {\rm [}b{\rm ]} + b^ - {\rm
[}a{\rm ]},$ получим формулу теоремы.

Из этой теоремы следует, что закон сохранения энергии
(\ref{(2.9)}) верен и в $D'_3 (R_4 ).$ Действительно, с учетом
(\ref{(3.2)}) и (\ref{(3.10)}) , имеем:
\[
\partial _t \hat W + div\hat P = \partial _t W + div\,P + (c\nu _4 [W]_F
+ [(\nu ,P)]_F ) =  - c\,{\mathop{\rm Re}\nolimits} (\hat j,\hat
A^* ).\] В силу (\ref{(2.9)}) скачок должен равняться нулю.

Пусть поле $A$ непрерывно по времени и терпит разрыв на
неподвижной поверхности $S \in R^3 $ (такие поля возникают при
дифракции волн на поверхностях с краем). В этом случае оно
удовлетворяет уравнению:
\begin{equation}\label{(3.11)}
- c^{ - 1} \partial _t \hat A -i\,rot\hat A = \hat j - in \times
[A]_S \delta _S(x).
\end{equation}
Это уравнение позволяет естественным образом ввести понятие
поверхностных токов.

О п р е д е л е н и е. Поверхностным ротором вектора $A$ на $S$
называется простой слой $ rot_S A = [A]_S \delta _S(x) \times n. $
Поверхностным током называется простой слой $j_S  = i\, rot_S A .$

З а м е ч а н и е. Введенные здесь определения поверхностных
дивергенции и ротора подобны введенным И.Е.Таммом [3], но не
совпадают с ними, а, фактически, соответствуют плотностям простых
слоев.

Таким образом поверхностные токи определяются через поверхностный
ротор, а поверхностные заряды -- через поверхностную дивергенцию.
При наличии таких поверхностей  из (\ref{(3.11)}) следует

\textit{Обобщенный закон сохранения заряда}:
\begin{equation}\label{(3.12)}
\partial _t \hat \rho  + div\hat j + i\,div\left( {
 [A]_S \delta_S(x) \times n} \right) = 0.
\end{equation}
В силу (\ref{(2.3)}) для объемных зарядов, отсюда следует

\textit{Закон сохранения поверхностных зарядов}:
\begin{equation}\label{(3.13)}
\partial _t \hat \rho _S  + div_S \hat j + i\,div\,rot_S A = 0
.
\end{equation}
Заметим, что в отличие от обычного ротора, дивергенция
поверхностного ротора не равна нулю. Это сингулярная  обобщенная
функция "типа двойного слоя".

{\bf 4.Тензор Грина.} При построении решений модифицированного
уравнения (\ref{(2.2)}) удобно пользоваться фундаментальным
решением этого уравнения  - тензором Грина.

О п р е д е л е н и е. Назовем $ U_k^l (x,t)$ тензором Грина
уравнения (\ref{(2.2)}) , если он  является его обобщенным
решением, соответствующим $j_k  = \delta _k^l \delta (x,t) $ (где
$\delta (x,t)$  --$\delta$-функция, $\delta _k^l $- символ
Кронекера) , и удовлетворяет условию излучения :
\begin{equation}\label{ (4.1)}
U_j^k = 0,\quad \left\| x \right\| > ct>0;\quad t<0.
\end{equation}
Для его построения воспользуемся преобразованием Фурье обобщенных
функций [2],  которое будем помечать черточкой: $\bar f(\xi
,\omega ) = F[f(x,t)]$.

Уравнение (\ref{(2.2)}) для функции Грина в пространстве
преобразований Фурье преобразуется в матричное алгебраическое вида
\begin{equation}\label{(4.2)}
L_{mj} (- i\xi ,- i\omega )\bar U_j^k = (i\omega c^{- 1} \delta
_{mj}+ e_{mjl} \xi _l )\bar U_j^k  = \delta _m^j ,\quad m,l,j,k =
1,2,3 .
\end{equation}
Здесь $(\xi ,\omega )$ -- переменные Фурье, соответствующие
$(x,t), \,\,\det {\bf L} = i\frac{\omega }{c}\left( {\left\| \xi
\right\|^2 - \frac{{\omega ^2 }}{{c^2 }}} \right). $

Разрешая (\ref{(4.2)}), найдем трансформанту Фурье функции Грина:
\begin{equation}\label{(4.3)}
\bar U_j^k  =  - \frac{{cQ_{jk} ( - i\xi , - i\omega )}}{{i\omega
\left( {(\omega /c)^2  - \left\| \xi  \right\|^2 } \right)}} ,
\end{equation}
где
$
Q_{km} ( - i\xi , - i\omega ) = \left\{ {\left( { - i\omega /c}
\right)^2 \delta _{km}  - \left( { - i\xi _k } \right)\left( { -
i\xi _m } \right) - i\,\omega c^{-1}\xi _j e_{kmj} } \right\}.
$

Введем функцию $ \bar \psi (\xi ,\omega ) = \left( {(\omega
/c)^2  - \left\| \xi \right\|^2 } \right)^{ - 1} $ --
преобразование Фурье функции Грина  волнового уравнения: $ \Delta
\psi  - c^{ - 2} \psi ,_{tt}  = \delta (x{\rm ,}t),\, $ которая
имеет вид:
\begin{equation}\label{ (4.5)}
\psi  =  - (4\pi R)^{ - 1} \delta (t - R/c) ,\quad R = \left\| x
\right\| .
\end{equation}
Будем назвать   $\psi$  \textit{волновой} функцией. Это простой
слой на  световом конусе - расширяющейся   сфере  радиуса $R =
ct\,\,(R = \left\| x \right\|$), определяющий линейный функционал
вида:
\[
(\psi ,\varphi ) = (4\pi )^{ - 1} \int\limits_{R^3 } {\left\| x
\right\|^{ - 1} \varphi (x,\left\| x \right\|/c)} dV(x),\forall
\varphi  \in D(R^4 ).
\]
Носителем этой функции является поверхность этого конуса.

Заметим, что знаменатель в представлении (\ref{(4.3)}), с учетом
носителя функций, является трансформантой Фурье   по времени
первообразной по $t$ волновой функции: $ \chi  = \psi \mathop
*\limits_t \theta (t) =  - (4\pi R)^{ - 1} \theta (ct - R),\quad
\partial _t \chi  = \psi $.  Здесь $\theta (.)$--функция
Хевисайда; переменная под звездочкой ($*$) -- знаком свертки --
означает неполную  свертку  только по этому аргументу. Носителем
$\chi$ является внутренность светового конуса. Пользуясь
свойствами преобразования Фурье свертки и производных, из
(\ref{(4.2)}) получим искомую функцию Грина:
\begin{equation}\label{(4.7)}
U_j^k (x,t) = c\,Q_{jk} (\partial _x ,\partial _t )\chi (R,t),
\end{equation}
Легко видеть, что условия излучения выполняются в силу свойств
волновой функции и ее первообразной. С учетом (\ref{(4.3)}),
результат сформулируем в виде теоремы.

Т е о р е м а 4.1. \textit{ Функция Грина модифицированных
уравнений Максвелла (\ref{(2.2)}) представима в виде}:
\[
U_j^k (x,t) = c^{ - 1} \delta _{jk} \partial _t \psi -
\partial _j \partial _k \chi  -i\, e_{jlk} \partial _l \psi.
 \]

Из ее свойств следует

Т е о р е м а 4.2. \textit{ Если токи $\hat j(x,t) \in D'_ +  (R^4
)$ и $ \mathop {supp}\limits_x \hat j$ ограничен, то обобщенное
решение уравнений (\ref{(2.2)}), соответствующее излучаемым и
затухающим на бесконечности волнам,  представимо в виде свертки}:
$ \hat A_m  = c\chi (R,t)*Q_{mk} (\partial _x ,\partial _t )\hat
j_k$, -- или
 \[
\hat A = c^{ - 1} \partial _t \psi *\hat j + c \nabla \psi *\hat
\rho -i\, [\nabla ,\psi *\hat j].
 \]

Д о к а з а т е л ь с т в о. Рассмотрим свертку $\hat A_m  = U_m^k
*\hat j_k.$ Легко показать, что это обобщенное решение уравнений
(\ref{(2.2)}), используя свойство дифференцирования свертки
\[
L_{lm} (\partial _x ,\partial _t )\hat A_m  = L_{lm} \left\{
{U_m^k *\hat j_k } \right\} = \left\{ {L_{lm} U_m^k }
\right\}*\hat j_k  = \delta _l^k \delta (x,t)*\hat j_k  = \hat
j_l.
\]
Далее, пользуясь  (\ref{(4.7)}), имеем
$
\hat A_m  = cQ_{mk} (\partial _x ,\partial _t )\chi (R,t)*\hat j_k
= c\chi (R,t)*Q_{mk} (\partial _x ,\partial _t )\hat j_k.
$
Либо, исходя из теоремы 4.1,
$
\hat A_m (x,t) = c^{ - 1} \psi (R,t)*\partial _t \hat j_m  -
c\,\chi (R,t)*\partial _m \partial _k \hat j_k  -i e_{mlk}
\partial _l \psi *\hat j_k.$
Отсюда, с учетом закона сохранения заряда (\ref{(2.3)}), получим
вторую формулу теоремы. Все свертки существуют в силу полу
ограниченности носителя по времени и ограниченности носителя
волновой функции по $x$. Излучение и затухание волн на
бесконечности следует из условий излучения для волной функции и из
свойства убывания плотности потенциала простого слоя на конусе как
$1/R$.

З а м е ч а н и е. В зависимости от свойства дифференцируемости
компонент свертки знак дифференцирования в формулах теоремы 4.2
можно перебрасывать на токи и заряды.

Т е о р е м а 4.3. \textit{Решение уравнения (\ref{(2.2)})
является решением волнового уравнения вида:}
\begin{equation}\label{(4.8)}
\Delta A - c^{ - 2} \frac{{\partial ^2 A}} {{\partial t^2 }} = -
i\,\textrm{\textrm{rot}}\,J + c\,\textrm{\textrm{grad}}\,\rho +
c^{ - 1}
\partial _t J .
\end{equation}

Д о к а з а т е л ь с т в о. Достаточно взять ротор в
(\ref{(2.2)}) , преобразовать двойной ротор с учетом равенства
\begin{equation}\label{(4.9)}
\textrm{rot}\,\textrm{rot} = \textrm{grad}\,\textrm{div}- \Delta ,
\end{equation}
а ротор под знаком производной по времени выразить через токи и
$\partial_tA$, еще раз используя (\ref{(2.2)}). Заметим, что
отсюда немедленно следует вторая формула теоремы 4.2.

П р и м е р  1 ("рождение"  заряда $q $ в точке $x=0$) . Пусть $
\rho  = q\delta (x)\theta (t)$. Тогда из закона сохранения заряда
получим: $ J =  \delta (t)q\,grad\left( {4\pi R} \right)^{ - 1}
$. Из теоремы 4.2 найдем поле:
\[
A = \,q\,\textrm{grad}\left\{ {c\left( {\theta (t)\mathop  *
\limits_t \psi } \right) +\,c^{ - 1} \partial _t \left( {\left(
{4\pi R} \right)^{ - 1} \mathop  * \limits_x \psi } \right)}
\right\}
\]
Это поле  можно записать в регулярном виде:
\[
4\pi A = \,q\,\textrm{grad}\,\left\{ { - \frac{{c\theta (ct -
R)}}{R} +
\partial _t \left( {\theta (ct - R) + \frac{{ct}}{R}\theta (R -
ct)} \right)} \right\} =
\]
\[
=\,q\,\textrm{grad}\left\{ { - c\,\theta (ct - R)/R -
 \left(1 - ct/R \right)\,\delta (t - R/c) -
 c\,\theta (R - ct)/R} \right\}
=  - q\,c\,\theta (t)\;\textrm{grad}\,(1/R)
\]
Т.е. поле совпадает с кулоновским полем статического заряда и
возникает вместе с зарядом.

{\bf 5. Постановка задачи Коши, единственность решений.}
Рассматривается A- поле, порождаемое нестационарным током с
плотностью из класса интегрируемых функций с компактным носителем
в $R^3$: $\mathop {supp}\limits_x j = \{ (x,t):\left\| x \right\|
< const,\,t > 0\} $ ограничен для любого $t$.

Начальные условия известны:
\begin{equation}\label{(5.1)}
A(x,0) = A^0 (x);\quad A^0  = 0,\quad \|x\| > const .
\end{equation}
Должны удовлетворяться условия излучения:
\begin{equation}\label{(5.2)}
A(x,t) = 0, \quad \|x\|  > const + ct,\quad\forall t \ge 0 .
\end{equation}

О п р е д е л е н и е. Назовем решение задачи Коши классическим,
если оно удовлетворяет условиям (\ref{(5.1)}) и (\ref{(5.2)}),
непрерывно  и дифференцируемо почти всюду кроме, быть может,
конечного числа волновых фронтов, на которых удовлетворяет
условиям на скачки (\ref{(3.6)}), а в области гладкости --
уравнениям (\ref{(2.2)}).

Запишем интегральную форму  закона сохранения энергии для решений
задачи Коши, в том числе для  ударных волн.

Пусть $S^{\rm  - } $ - произвольное открытое множество в $R^3 $,
ограниченное гладкой замкнутой поверхностью $ S, \,  n$- единичный
вектор нормали к $S$.

Т е о р е м а 5.1. \textit{Если $A(x,t)$ -- классическое решение
задачи Коши, то}
\[
\int\limits_{s^ -  } {(W(x,t) - W(x,0))} dV(x) + \int\limits_0^t
{dt\int\limits_s {(n,P)dS(x)} }  =  - {\mathop{\rm Re}\nolimits}
\int\limits_0^t {dt\int\limits_{s^ -  } {(j,A^* )} } dV(x),
\]
\[
\int\limits_{R^3 } {(W(x,t) - W(x,0))dV(x)}  =  - {\mathop{\rm
Re}\nolimits} \int\limits_0^t {dt} \int\limits_{R^3 } {(j,A^*
)dV(x)},
\]
где $dV(x)=dx_1 dx_2 dx_3$.

Д о к а з а т е л ь с т в о.    Проинтегрируем (\ref{(2.9)}) в
$R^4$ по области $ D_{t}^ -   = S^ -   \times (0,t)$  и, пользуясь
формулой Остроградского-Гаусса, с учетом возможных разрывов
решений на фронтах, получим:
\begin{equation}\label{(5.3)}
\int\limits_{S^ -  } {(W(x,t) - W(x,0))} dV(x) + \int\limits_0^t
{dt\int\limits_S {(n,P)dS(x)} } + \int\limits_{F_{} } {(\nu _j
{\rm [}P_j {\rm ]}}  + \nu _4 {\rm [}W(x,t){\rm ]})dF =
\end{equation}
\[
=- {\mathop{\rm Re}\nolimits} \int\limits_0^t {dt\int\limits_{S^ -
} {(j,A^* )} } dV(x)  ,
\]
 где $dF$- дифференциал площади
характеристической поверхности $F$, в квадратных скобках стоят
скачки соответствующих выражений на $F$.

Скачок на волновых фронтах подынтегрального выражения в третьем
интеграле, в силу условия на фронтах (\ref{(3.10)}), равен нулю. В
результате получаем первую формулу теоремы. Вторая формула следует
из нее же с учетом условия на бесконечности (\ref{(5.2)}).

Ясно, что условия ограниченности на токи можно ослабить,
достаточно, чтобы интегралы, их содержащие, существовали.

В силу положительной определенности $W$, из теоремы 5.1 получим

С л е д с т в и е. Классическое решение задачи Коши единственно.

Д о к а з а т е л ь с т в о от противного. Пусть есть два решения,
тогда  интеграл энергии для их разности А имеет вид:
$\int\limits_{R^3 } {W(x,t)dV(x)}  = 0.$ В силу положительной
определенности $W$, получим: $A = 0$. Следовательно эти решения
совпадают.

{\bf 6. Задача Коши в $ D_3 ^\prime  (R^4_+ )$ и ее решение.}
Пусть A - решение  задачи Коши. Доопределим его на $R^4_+ $ , а
именно, введем обобщенную функцию $ \hat A = A\theta (t)$.
Аналогично строим продолжение на $R^4 $для токов $\hat j$, и будем
их рассматривать как обобщенные функции на $D_3 ^\prime  (R^4 )$.
Подействуем на  $A$ оператором $L$. С учетом (\ref{(2.2)}) ,
получим
\begin{equation}\label{(6.1)}
L_{kl} (\partial _x ,\partial _t )\,\hat A_l  = \hat j_k  - c^{ -
1}\, A_k^0 \delta (t).
\end{equation}
Сравнивая вторые слагаемые в правой части с (\ref{(2.2)}), видим,
что начальные условия работают как импульсные объемные токи.

Обобщенное решение (\ref{(6.1)}) строится с помощью функции Грина
$U(x,t)$ (см. теорема 4.1).

Т е о р е м а 6.1. \textit{ Если $A_0  \in D_3' (R^3 ),\,\hat j
\in D_3' (R_{^ + }^4 ),$ то
\[
\hat A = c^{ - 1} \partial _t \psi *\hat j +c\, \nabla \psi *\hat
\rho  -i\, [\nabla ,\psi *\hat j] - c ^{ - 2} \partial _t (\psi
\mathop  * \limits_x A_{}^0 ) -c\, \nabla (\psi \mathop  *
\limits_x \hat \rho ^0 ) +i [\nabla ,\psi \mathop *\limits_x A^0
],
\]
\[
\hat \rho  = \rho _0 \theta (t) - div\,(\hat j\mathop *\limits_t
\theta (t)),\quad \hat \rho _0  =c ^{ - 1} div\hat A_0.
\]}
Д о к а з а т е л ь с т в о. Используя свойство функции Грина,
представим решение в виде свертки: $ \hat A_l  = U_l^k  * \,\hat
j_k  - c^{ - 1} U_l^k \mathop  * \limits_x A_k^0$ (здесь знак
"$\mathop *\limits_x $ " указывает неполную свертку только по
$x$). Далее  покомпонентно берем все свертки, которые существуют в
силу полуограниченности по $t$ носителей входящих в них функций и
ограниченности по$x$ при фиксированном $t$ носителя волновой
функции и ее первообразной по времени. С учетом формулы заряда
(\ref{(2.3)}) и закона сохранения заряда  в результате получим
утверждение теоремы.

Для того чтобы записать решение в интегральном виде, следует
воспользоваться формулами для сверток с  $\psi (x,t)$ ). Запись
решения в сверточном виде очень удобна для перехода к их
интегральным представлениям, т.к. в зависимости от свойств
гладкости входящих в свертки функций, можно перебрасывать
дифференцирование на ее составляющие, допускающие обычное
дифференцирование, что позволяет вводить знак дифференцирования
под интегралы с переменной областью интегрирования. Последнее
является характерной особенностью нестационарных задач и
достаточно затруднительно при применении классических методов.

П р и м е р 2 (распространение ударных волн). Предположим, что
только в области $S^-$, ограниченной замкнутой поверхностью $S$,
которая является фронтом ударной волны, начальное поле отлично от
нуля  и $A^0  \in C^1 (S^ -  )$.  Пусть токи  отсутствуют $ (j =
0)$ и начальный заряд $\rho _0  = div\,A^0 = 0$. Найдем поле в
$R^3$ при $t>0$.

Введем характеристическую функцию $H_S^ -  (x)$ множества $S^ - $,
равную 1 на $S^ -  $ и 0 вне его замыкания. В силу условий на
фронтах $ \hat \rho _0  =  \rho _0  + \rho _S  = 0$. Поэтому,
согласно теореме 6.1, в этом случае $\hat A =  - c^{ - 2} \partial
_t (\psi \mathop  * \limits_x A_{}^0 H_S^ -  (x))+i\,c ^{ -1}
rot\,(\psi \mathop *\limits_x A^0 H_S^ - (x))$. Последнее можно
записать в интегральном виде:
\[
4\pi\,c A =  -c ^{ -1} \partial _t ((ct)^{ - 1} \int\limits_{r =
ct} {A^0 (y)H_S^ -  (y)dS(y)})  + i(ct)^{-1}\,\int\limits_{r = ct}
{rot\,A^0 } (y)dS(y),\quad r = \left\| {x - y} \right\|.
\]
Эта формула описывает вихревое  ЭМ поле при распространении
ударных волн.

Рассмотрим еще некоторые, важные с точки зрения приложений, типы
ЭМ полей.

\textbf{7.\,Монохроматическое А-поле.} \,В радиофизике и
радиотехнике изучаются и используются процессы,  когда действующие
токи и соответственно ЭМ-поля имеют гармоническую зависимость от
времени с частотой $\omega ,\,\omega  > 0$ :$ J = J(x)\exp ( -
i\omega t), \,   A = A(x)\exp ( - i\omega t) $. В этом случае из
теоремы 4.3 следует

Т е о р е м а  7.1. \textit{Вектор комплексной амплитуды А
удовлетворяет уравнению $\Delta A + k^2 A = - ik\,J -
i\,\textrm{rot}\,J + c\,\textrm{grad}\,\rho, $ и если $J\in
D_3'(R^3 )$ и $ supp\, J(x)$ ограничен, то}
\[
4\pi A = ik\frac{{e^{ikR} }} {R}*J + i\,\textrm{rot}\left(
{\frac{{e^{ikR} }} {R}*J} \right) - c\,\textrm{grad}\,\left(
{\frac{{e^{ikR} }} {R}*\rho } \right) ,\quad \rho  =  - i\omega ^{
- 1} \textrm{div} \,J.
\]
\textit{Это решение единственно в классе решений, удовлетворяющих
условиям излучения Зоммерфельда на бесконечности }[2].

\textbf{8.\,Стационарное А-поле. } \,Пусть A-поле не зависит от
времени. Тогда из теоремы 4.2 получим для А уравнение Пуассона:
\begin{equation}\label{(8.1)}
\Delta A =  - i\,\textrm{rot}\,J + c\,\textrm{grad}\,\rho ,
\end{equation}
решение которого имеет вид  ньютоновского потенциала [2]. А
именно, верна следующая

Т е о р е м а  8.1. \textit{Если $J \in D_3' (R^3 )$  и
$supp\,J(x)$ ограничен, то решение (\ref{(8.1)}), стремящееся к
нулю на бесконечности, единственно и  представимо в виде свертки}
\begin{equation}\label{(8.2)}
A = - c\,\textrm{grad}\,\left( {\frac{1} {{4\pi R}}*\rho } \right)
+ i\,\textrm{rot}\,\left( {\frac{1} {{4\pi R}}*J} \right).
\end{equation}

Если ввести скалярный и векторный потенциалы А-поля: $c ^{ - 1} A
= \textrm{grad}\Phi  + i\,\textrm{rot}\Psi ,\,\, \textrm{div}\Psi
= 0, $ то, подставляя в (\ref{(8.1)}), получим также уравнения
Пуассона: $ \Delta \Phi  = \rho ,\quad\quad \Delta \Psi = - c^{ -
1} J,$ откуда следует представление потенциалов стационарного ЭМ
поля через ньютоновские потенциалы:
\begin{equation}\label{(8.3)}
\Phi  =  - \frac{1} {{4\pi R}} * \rho , \quad\quad \Psi =
\frac{1}{{4\pi c\,R}} * J,
\end{equation}
что также следует и из (\ref{(8.2)}).

П р и м е р 3. Рассмотрим комплексные токи в шаре, вращающиеся
вокруг оси $X_3$ с постоянной угловой скоростью $\omega$:
\[
J = \rho (x)\left[ {\varpi  \times x} \right] ,\quad \varpi  =
({\rm 0}{\rm ,\,0}{\rm ,}\,\omega ) ; \quad \rho (x) =
0,\quad\left\| x \right\|
> a.
\]
Тогда
\[
\Phi  =  - \frac{1}{{4\pi }}\int\limits_{\left\| y \right\| < a}
{\frac{{\rho ^{} (y)}}{{\left\| {x - y} \right\|}}} dV(y) ,\,\,\,
\Psi  =  - \frac{\omega }{{4\pi }}\int\limits_{\left\| y \right\|
< a} {\frac{{\rho (y)}}{{\left\| {x - y} \right\|}}} \left(
{\begin{array}{*{20}c}
   { - y_2 }  \\
   {y_1 }  \\
   0  \\
\end{array}} \right)dV(y)
\]
Если плотность токов - константа: $\rho  = \rho _0 $, то интегралы
можно вычислить аналитически. Например,
\[
\Phi  =  - \frac{{\rho _0 }}{{2}}\int\limits_{R' < a} {R'^2
dR'\int\limits_0^\pi  {} \frac{{\sin \vartheta d\vartheta
}}{{\sqrt {R^2  + R'^2  - 2RR'\cos \vartheta } }}}  = \frac{{\rho
_0 }}{{2R}}\int\limits_{R' < a} {R'dR'\int\limits_{ - R'}^{R'}
{d\sqrt {R^2  + R'^2  - 2Rz} } }  =
\]
\[
= \frac{{\rho _0 }}{{2R}}\int\limits_{R' < a} {R'} \left( {\left|
{R - R'} \right| - R - R'} \right)dR' =  - \{
{\begin{array}{*{20}c}
   {\rho _0 a^3 /3R,\quad \quad \quad \quad R \geq a}  \\
   {0,5\rho _0 (a^2  - R^2 /3),\;\;R \leq a}  \\
\end{array}} .
\]
Отсюда следует известный результат, что вне шара потенциальная
часть поля совпадает с полем точечного заряда $q  = q_ 0\,\delta
(x)$, равного заряду шара $q _ 0 = \frac{{4\pi a^3 \rho _0 }}{3}$,
сосредоточенного в точке $x=0$. Она имеет следующий вид:
\[
\textrm{grad}\,\Phi  = \left\{ {\frac{{\rho _0 x}}{3},\;R \leq
a;\;\;\frac{{\rho _0 a^3 }}{{3R^2 }}\,\frac{x}{R},\;R \geq a}
\right\}, -
\]
и максимальна на поверхности шара, а вне его убывает как $1/R^2$.

Аналогично находим векторный потенциал:
\[
\Psi _2  =  - \frac{{\rho _0 }}{{4\pi R}}\int\limits_{R' < a}
{R'^3 dR'\int\limits_0^\pi  {\sin \vartheta d\vartheta
\int\limits_0^{2\pi } {} } \frac{{\sqrt {x_1^2  + x_2^2 } \sin
\vartheta \cos \varphi  + x_1 \cos \vartheta }}{{\sqrt {R^2  +
R'^2  - 2RR'\cos \vartheta } }}d\varphi }  =
\]
\[
= \frac{{\rho _0 a^3 x_1 }}{{3\pi R}}\left\{
   {a^2 /4R^2 ,\,\, R \geq a} ;\quad
   {aR - 3R^2 /4,\,\, R \leq a}
 \right\}\]
 Также определяется
\[
\Psi _1  =  - \frac{{\rho _0 \omega x_2 }}{{3\pi R}}\{ {\frac{{a^4
}}{{4R^2 }}} ,\,\, R \geq a;\quad {R(a - \frac{3}{4}R),\,\,R \leq
a} \}.
\]
Далее находим вихревую составляющую А-поля
\[
i\,\textrm{rot}\,\Psi  =   - \frac{{i\,\rho _0 \omega }}{{3\pi
}}\left\{ {\begin{array}{*{20}c} {\frac{{a^4 }}{{R^3 }}\left(
{\frac{{3x_1 x_3 }}{{4R^2 }},\, \frac{{3x_2 x_3 }}{{4R^2
}},\,\frac{1}{2} - \frac{{3r^2 }}{{4R^2 }}} \right),\quad R \geq
a} \,\, \\{\left( {\frac{{3x_1 x_3 }}{{4R}},\,\frac{{3x_2 x_3 }}
{{4R}},\,2a - \frac{3}{2}R(1 + \frac{{r^2 }}{{2R^2 }})}
\right),\quad R \leq a}  \\
\end{array}}\right. ,\,\,\,
r = \sqrt {x_1^2  + x_2^2 }. \]

Заметим, что как потенциалы , так и  поле $A$  непрерывны при
любых $x$, в том числе и при $R=a$.

{ \bf Заключение}. Обозначим <a> размерность $a$. Согласно
определению квадрат модуля $A$ равен плотности энергии, т.е
$<\|A\|^2>=$(дж/м$^3)$. Поэтому это А-поле можно назвать
энергетическим. Возможность описания ЭМ поля через один  вектор
убедительно свидетельствует о неразрывной связи между
электричеством и магнетизмом, которые просто являются физическим
проявлением комплексного A-поля. Единственной константой этого
поля является скорость ЭМ волн $c$, а не две ($\varepsilon,\mu$ )
ЭМ поля. Заметим, что многие формулы электродинамики можно
упростить, если использовать A-вектор.

\vspace{3mm}
\begin{center}
\bf {ЛИТЕРАТУРА}
\end{center}
{\footnotesize  

\makebox[2em][r]{1. }{\it Ахиезер А.И., Берестецкий В.Б.}Квантовая электродинамика. М., 1981.

\makebox[2em][r]{2. }  {\it Алексеева Л.А.}//Журнал
вычислительной математики и математической физики. Т.35(2202),
№1,С.125-137.

\makebox[2em][r]{3. }  {\it  Владимиров В.С.} Уравнения
математической физики. М. 1981.

\makebox[2em][r]{4. }  {\it  Тамм И.Е.} Основы теории
электричества. М.: Наука, 1989.}

\end{document}